# Heterogeneous anomalous diffusion of virus in cytoplasm of a living cell

Yuichi Itto

*Department of Physical Engineering, Mie University, Mie 514-8507, Japan*

**Abstract.**  The infection pathway of virus in cytoplasm of a living cell is studied from the viewpoint of diffusion theory. The cytoplasm plays a role of a medium for stochastic motion of the virus contained in the endosome as well as the free virus. It is experimentally known that the exponent of anomalous diffusion fluctuates in localized areas of the cytoplasm. Here, generalizing fractional kinetic theory, such fluctuations are described in terms of the exponent locally distributed over the cytoplasm, and a theoretical proposition is presented for its statistical form. The proposed fluctuations may be examined in an experiment of heterogeneous diffusion in the infection pathway.

In recent years, an exotic phenomenon has experimentally been observed by making use of the technique of real-time single-molecule imaging in the infection pathway of adeno-associated viruses in cytoplasm of a living *HeLa* cell [1,2]. (Here, the adeno-associated virus is a small virus particle, and the *HeLa* cell is a line of human epithelial cells.) In each experiment, the virus solution of low concentrations, in which the virus is labeled with fluorescent dye molecule, was added to a culture medium of the living cells. Then, the trajectories of the fluorescent viruses in the cytoplasm were observed. The experiments show that the adeno-associated virus exhibits stochastic motion inside the cytoplasm in two different forms: one is in the free form and the other is the form being contained in the endosome (i.e., a spherical vesicle, see Fig. 1).

Let $\overline{x^2}$ be the mean square displacement in stochastic motion. In general, it scales for large elapsed time, $t$, as

$$\overline{x^2} \sim t^{\alpha}. \tag{1}$$

Normal diffusion reads $\alpha = 1$, whereas $0 < \alpha < 1$ ($\alpha > 1$) corresponds to subdiffusion (superdiffusion). The result of experimental observation shows that the trajectory of the virus exhibits not only $\alpha = 1$ but also $0 < \alpha < 1$ in the form of Eq.

(1). However, what is truly remarkable is the fact [1] that, in the case of subdiffusion, $\alpha$ fluctuates between 0.5 and 0.9, depending on localized areas of the cytoplasm. This manifests the *heterogeneous* structure of the cytoplasm as a medium for stochastic motion. It is noted [1] that this heterogeneity is not due to the forms of existence of the virus (i.e., being free or contained in the endosome). Thus, this phenomenon is in marked contrast to traditional anomalous diffusion [3] discussed for physical systems, such as particle motion in turbulent flow [4], charge carrier transport in amorphous solids [5], the flow of contaminated vortex in fluid [6], chaotic dynamics [7], porous glasses [8], and so on.

The experimental result mentioned above poses a novel interesting problem for the physics of diffusion. On the other hand, in biology, it is essential to understand the virus infection process for both designing antiviral drug and developing efficient gene therapy vectors. It is therefore of obvious importance to investigate the virus infection pathway from the physical viewpoint.

In this paper, we study the infection pathway of the adeno-associated virus in the cytoplasm of the living *HeLa* cell by generalizing traditional theory of anomalous diffusion. We regard the cytoplasm as a medium for stochastic motions of both the free virus and the virus contained in the endosome. Then, we imaginarily divide the medium into many small blocks. In other words, a block is identified with a localized area of the cytoplasm. This procedure seems to be necessary when the infection pathway of the virus in the entire cytoplasm is considered. The mean square displacement of the virus does not always show normal diffusion and/or subdiffusion with a fixed exponent, since the virus in a given localized area moves to neighboring ones before reaching the nucleus of the cell [1]. Thus, the exponent, $\alpha$, in Eq. (1) locally fluctuates from one block to another in the cytoplasm. Furthermore, we consider that this fluctuation varies slowly over a period of time, which is much longer than the time scale of the stochastic motion of virus in a localized area of the cytoplasm. It is therefore assumed that there is a large time-scale separation in the infection pathway. For the virus in each block, we apply fractional kinetic theory, which generalizes Einstein's approach to Brownian motion [9]. Generalizing traditional fractional kinetic theory, we describe the local fluctuations of the exponent, $\alpha$. We propose the statistical form of fluctuations from the experimental data. Then, we show that the proposed form of fluctuations can be derived by the maximum entropy principle [10].

Let us start our discussion with the motion of the virus in a one-dimensional block (i.e., a segment). To describe it, we consider the following evolution equation based on the scheme of continuous-time random walks [11]:

$$f(x,t)\,dx = dx \int_{-\infty}^{\infty} d\Delta \int_{0}^{t} d\tau\, f(x+\Delta, t-\tau)\, \phi_{\tau}(\Delta)\, \psi(\tau) + \delta(x)\, R(t)\, dx, \qquad (2)$$

where $f(x,t)\,dx$ is the probability of finding a virus in the interval $[x, x+dx]$ at time $t$. The first term on the right-hand side describes all of possible probabilities that the virus moves into the interval from outside or stays in the interval. Then, the second term is a partial source guaranteeing the initial condition, $f(x,0) = \delta(x)$, from which the condition, $R(0) = 1$, follows. $\phi_{\tau}(\Delta)$ is the normalized probability density distribution for a displacement, $\Delta$, in a finite time step, $\tau$. This distribution is sharply peaked at $\Delta = 0$ and fulfills the condition, $\phi_{\tau}(\Delta) = \phi_{\tau}(-\Delta)$. $\tau$ in Eq. (2) is treated as a random variable following the normalized probability density, $\psi(\tau)$, which satisfies $\psi(0) = 0$. $R(t)$ describes a time-dependent partial source and is connected to $\psi(\tau)$ through the relation: $R(t) = 1 - \int_{0}^{t} d\tau\, \psi(\tau)$, which comes from the normalization condition for $f(x,t)$. In particular, in the deterministic case, $\psi(\tau) = \delta(\tau - \tau_0)$, Eq. (2) becomes reduced to the basic equation in Einstein's approach to Brownian motion [9] after the replacement, $t \to t + \tau_0$.

Now, an origin of subdiffusion found in the experiments may be not in $\phi_{\tau}(\Delta)$ but in $\psi(\tau)$. Accordingly, we assume in what follows that $\phi_{\tau}(\Delta)$ is actually independent of time steps: $\phi_\tau(\Delta) = \phi(\Delta)$. To see how Eq. (2) leads to subdiffusion, we here employ the Laplace transform of Eq. (2) with respect to time:

$$\widetilde{f}(x,u) = \int_{-\infty}^{\infty} d\Delta \widetilde{f}(x+\Delta,u)\phi(\Delta)\widetilde{\psi}(u) + \delta(x)\frac{1-\widetilde{\psi}(u)}{u}, \qquad (3)$$

where $\widetilde{f}(x,u)$ and $\widetilde{\psi}(u)$ are the Laplace transforms of $f(x,t)$ and $\psi(\tau)$, respectively, provided that $\mathcal{L}(g)(u) = \widetilde{g}(u) = \int_0^\infty dt' g(t')e^{-ut'}$. Then, we require $\widetilde{\psi}(u)$ to have the following form:

$$\widetilde{\psi}(u) = \widetilde{\psi}_\alpha(u) \sim 1 - (su)^\alpha. \qquad (4)$$

Here, $s$ is a characteristic constant with the dimension of time. This characteristic time is an indicative one, at which the virus is displaced. We also impose the condition that $\psi(\tau)$ has the divergent first moment, so that the exponent $\alpha$ is in the interval $(0,1)$. Eq. (4) implies that $\psi(\tau)$ decays as a power law like, $\psi(\tau) \sim s^\alpha / \tau^{1+\alpha}$, for the time step $\tau$ longer than $s$. Later, we shall show how traditional fractional kinetic theory [12] is derived from Eqs. (3) and (4). The mean square displacement of the virus turns out to have the form in Eq. (1), reproducing the behavior observed in a localized area of the cytoplasm. In particular, the case of normal diffusion is realized in the limit, $\alpha \to 1$, of the present theory.

The virus moves from one block to another, and in such a process the exponent $\alpha$ locally fluctuates over the cytoplasm. We shall therefore develop a generalized fractional kinetic theory, in which this fluctuation is incorporated. To do so, it is essential to clarify the statistical property of the fluctuations. According to the experiment [1], the trajectories of 104 viruses are analyzed. 53 trajectories among them exhibit $\alpha = 1$ in Eq. (1), and other 51 show $\alpha$ varying between 0.5 and 0.9. Besides this fact, no further information is available about the weights of $\alpha \in (0.5, 0.9)$. Although $\alpha$ for the virus contained in the endosome might be different from that for the free virus in general, we here assume that the exponents found in both the free and endosomal forms differ from each other only slightly. The virus tends to reach the nucleus of the cell. Due to this tendency, the exponent near $\alpha = 0$ may seldom be realized. On the other hand, normal diffusion is often to be the case. From these considerations, we propose a Poisson-like form of fluctuations:

$$P(\alpha) \sim e^{\lambda \alpha} \qquad (5)$$

with a positive constant, $\lambda$. We note that this distribution can be derived from the maximum entropy principle [13] with the Shannon entropy, $S[P] = -\int_0^1 d\alpha\, P(\alpha) \ln P(\alpha)$, under the constraints on the normalization condition, $\int_0^1 d\alpha P(\alpha) = 1$, and the expectation value of $\alpha$, $\int_0^1 d\alpha\, \alpha P(\alpha) = \overline{\alpha}$. In this case, $\lambda$ is a positive Lagrange multiplier associated with the constraint on the expectation value of $\alpha$.

Next, we formulate the generalized fractional kinetic theory based on the distribution in Eq. (5). As mentioned earlier, $\alpha$ slowly varies locally but is approximately constant while the virus moves through the blocks in the entire cytoplasm. So, the effective distribution of the time step in the Laplace space appears a superposition of $\widetilde{\psi}_\alpha(u)$ with respect to $P(\alpha)$:

$$\widetilde{\psi}(u) = \int_0^1 d\alpha\, P(\alpha) \widetilde{\psi}_\alpha(u). \tag{6}$$

Substituting Eq. (6) into Eq. (3), expanding $\widetilde{f}$ up to the second order of $\Delta$, and neglecting the term $<\Delta^2> \int_0^1 d\alpha P(\alpha)(su)^\alpha$ ($u$ being small in the long time behavior) with $<\Delta^2> \equiv \int_{-\infty}^\infty d\Delta\, \Delta^2 \phi(\Delta)$, we find

$$\widetilde{f}(x,u) = \frac{<\Delta^2>}{2\int_0^1 d\alpha P(\alpha)(su)^\alpha} \frac{\partial^2 \widetilde{f}(x,u)}{\partial x^2} + \delta(x)\frac{1}{u}. \qquad (7)$$

Equivalently, performing the inverse Laplace transform of Eq. (7), we obtain the following generalized fractional diffusion equation:

$$\int_0^1 d\alpha\, P(\alpha)\, s^{\alpha-1}\, {}_0\mathcal{D}_t^{-(1-\alpha)}\, \frac{\partial f(x,t)}{\partial t} = D\frac{\partial^2 f(x,t)}{\partial x^2}, \qquad (8)$$

where $D$ is the diffusion constant calculated as $D = <\Delta^2>/(2s)$ and a mathematical fact of fractional operator [12], $\mathcal{L}({}_0\mathcal{D}_t^{-\alpha} g(x,t))(u) = u^{-\alpha}\widetilde{g}(x,u),$ is used. For the virus in a given local block, the distribution of fluctuations is taken to be $P(\alpha) = \delta(\alpha - \alpha_0)$ in Eq. (8). This leads to the traditional fractional kinetic theory [12] after applying the operator, $(\partial/\partial t)\, {}_0\mathcal{D}_t^{-\alpha_0}$, to Eq. (8), as mentioned earlier.

Eq. (8) is the main result of our theory describing the infection pathway of the virus in the entire cytoplasm.

It should be recalled that the adeno-associated virus is in the cytoplasm, freely as well as in the form being contained in the endosome. Theoretically, the discrimination between them could result in different values of the diffusion constant $D$, as performed in the experiments [1,2].

In conclusion, we have studied the infection pathway of an adeno-associated virus in cytoplasm of a living *HeLa* cell from the viewpoint of diffusion theory. We have regarded the cytoplasm as a medium for stochastic motions of both the free virus and the virus contained in the endosom. Dividing the cytoplasm into many small virtual blocks, we have described the local fluctuations of the exponent of subdiffusion of the virus. We have assumed slowness of variation of the exponent and proposed the form of its statistical distribution based on the maximal entropy principle. Then, we have formulated the generalized fractional kinetic theory by introducing heterogeneity of diffusion. It is of extreme interest to examine if the proposed fluctuation distribution in Eq. (5) is realized in an experiment of heterogeneous diffusion in the infection pathway under the same experimental conditions.

**ACKNOWLEDGEMENTS**

The author thanks Professor Abe for his comments and reading of the manuscript. He also thanks Dr Y. Ookura at the University of Tsukuba for providing him with the relevant references.

---

[1] G. Seisenberger, M.U. Ried, T. Endreß, H. Büning, M. Hallek, and C. Bräuchle, Science **294**, 1929 (2001).

[2] C. Bräuchle, G. Seisenberger, T. Endreß, M.U. Ried, H. Büning, and M. Hallek, ChemPhysChem **3**, 299 (2002).

[3] J.-P. Bouchaud and A. Georges, Phys. Rep. **195**, 127 (1990).

[4] L.F. Richardson, Proc. Roy. Soc. London, A **110**, 709 (1926).

[5] H. Scher and E.W. Montroll, Phys. Rev. B **12**, 2455 (1975).

[6] O. Cardoso and P. Tabeling, Europhys. Lett. **7**, 225 (1988).

[7] M.F. Shlesinger, G.M. Zaslavsky, and J. Klafter, Nature (London) **363**, 31 (1993).

[8] S. Stapf, R. Kimmich, and R.-O. Seitter, Phys. Rev. Lett. **75**, 2855 (1995).

[9] A. Einstein, Ann. Phys. (Leipzig) **17**, 549 (1905) (English transl. Investigations on the Theory of the Brownian Movement, Dover, New York, 1956).

[10] E.T. Jaynes, Phys. Rev. **106**, 620 (1957).

[11] E.W. Montroll, and G.H. Weiss, J. Math. Phys. **6**, 167 (1965).

[12] R. Metzler and J. Klafter, Phys. Rep. **339**, 1 (2000).

[13] Y. Itto, in preparation.

# Figure Caption

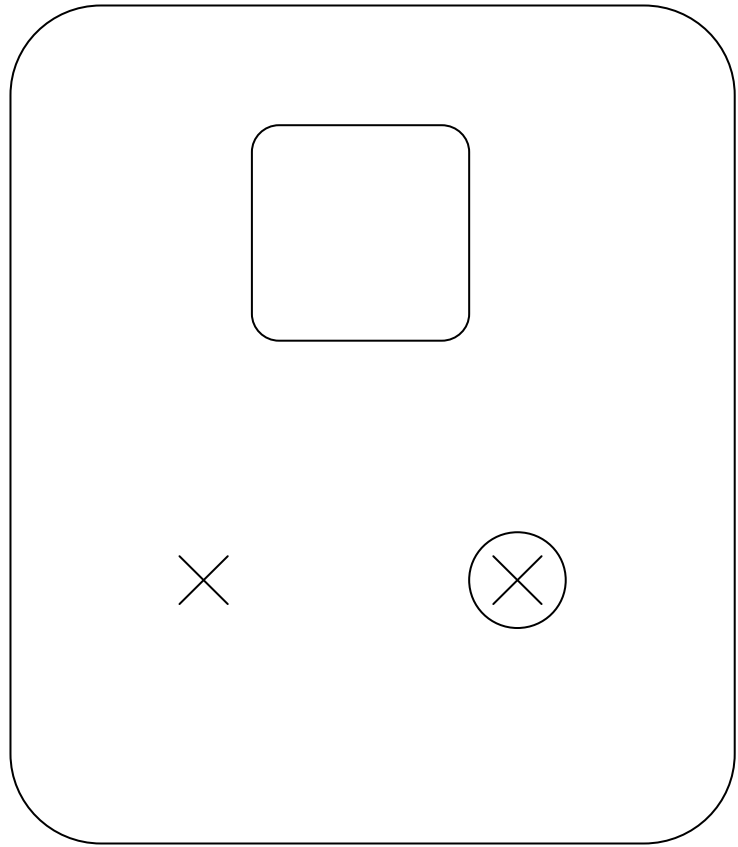

# Figure 1.

Figure 1.   Two of the adeno-associated viruses in the cytoplasm of the living *HeLa* cell. One is free, and the other is contained in the endosome. The cross stands for the virus, whereas the circle depicts the endosome. The large and small boxes represent the cell and nucleus, respectively.